\documentclass[final,5p,times,twocolumn,author-year]{elsarticle}
\setcitestyle{square}
\usepackage{amssymb}
\usepackage{amsmath}
\usepackage{cancel}
\usepackage{float}
\usepackage[section]{placeins}
\usepackage[table,xcdraw]{xcolor}
\usepackage{adjustbox}
\usepackage{multirow}
\usepackage{hyperref}
\journal{Acta Astronautica}

\begin{document}

\begin{frontmatter}



\title{Investigation of minimum frame rate for low-latency planetary surface teleoperations}


\author[label1]{Benjamin Mellinkoff}
\author[label1]{Matthew Spydell}
\author[label2]{Wendy Bailey}
\author[label1]{Jack O. Burns}
\address[label1]{Center for Astrophysics and Space Astronomy, 593 UCB, University of Colorado Boulder, Boulder, CO 80309, USA}
\address[label2]{Engineering Management Program, University of Colorado Boulder, Boulder, CO 80309, USA}

\begin{abstract}
The Global Exploration Roadmap indicates the need for increased human exploration of under-sampled regions of our solar system in order to make new scientific discoveries \cite{GER}. The high costs and dangers of sending humans deeper into our solar system necessitates the use of human-robotic partnerships, especially in transitioning from low-Earth orbit to deep-space operations. Low-latency planetary surface exploration is an example of a human-robotic partnership that provides an exciting option for effective, low-cost exploration of our solar system. However, low-latency telerobotic exploration is a new concept for space exploration and needs to be tested for its limits and effectiveness. This paper focuses on a human operator's ability to identify exploration targets in an unfamiliar environment using real-time low-latency telerobotics under various frame rate conditions. This relationship was investigated using a Telerobotic Simulation System (TSS). The frame rates were varied and the order of the exploration tasks were randomized for each operator. The rover operated at peak speeds of one meter per second with a video stream resolution of 640x480 and colorscale of 24 bits. The results from this experiment indicate that 5 frames per second is the minimum necessary frame rate for effective exploration.
\end{abstract}


\end{frontmatter}



%%%%%%%%%%%%%     INTRO     %%%%%%%%%%%%%%
\section{Introduction}

\subsection{Low-Latency Surface Telerobotics}
The use of low-latency surface telerobotics will become a keystone of future human space missions. This mode of exploration will become pivotal because of its ability to expedite scientific research in the areas of astrophysics and planetary science. The purpose behind low-latency surface telerobotics is to combine the deployability of robotics with human ingenuity to enhance scientific exploration in space that produces rapid meaningful results.\par 
In previous space missions scientific exploration was primarily accomplished using either robotics or human astronauts, rarely were these two approaches fully integrated with each other. The use of robotics is advantageous for exploring hostile and distant environments in the solar system; the drawback is the dramatic limitation in situational awareness. The current rovers on Mars provide a good example for high-latency telerobotic missions. Round-trip latency times range from 8.6 minutes to 42 minutes making it impossible for real-time control via humans \cite{HERRO}. This forces most of the computing and decision-making to be done strictly on board the rover, as opposed to offloading the processing to more powerful systems. Therefore, the operating speeds of these high-latency rovers are considerably slower than a comparable system with increased human involvement.\par
On the other hand, the scientific exploration that occurred during the Apollo missions resulted primarily from a physical human presence on the lunar surface. The use of human astronauts is advantageous when considering the time efficiency of the scientific exploration accomplished; however, the drawback is that human surface missions are high risk and costly. Combining humans and robots to be used in teleoperations yields the benefits from both methods of exploration while removing many of the drawbacks from using humans or robots independently.

\subsection{Telerobotic Applications on Mars}
In the initial stages of exploring Mars, low-latency surface teleoperated vehicles will be controlled by astronauts in orbit. This will allow for real-time communication with the vehicles on the surface of Mars that will drastically improve the amount of scientific exploration accomplished in a given amount of time. 
Low-latency surface telerobotics also allow for a virtual ``human presence" on the Martian surface without physically landing \cite{humanPres}. The first human missions to Mars will likely be orbital missions, so the astronauts will use a fleet of teleoperated vehicles to explore and begin establishing the infrastructure required for future human surface missions.

\subsection{Telerobotic Applications on the Moon}
\begin{figure*}[h]
\centering
\includegraphics[width=0.98\textwidth]{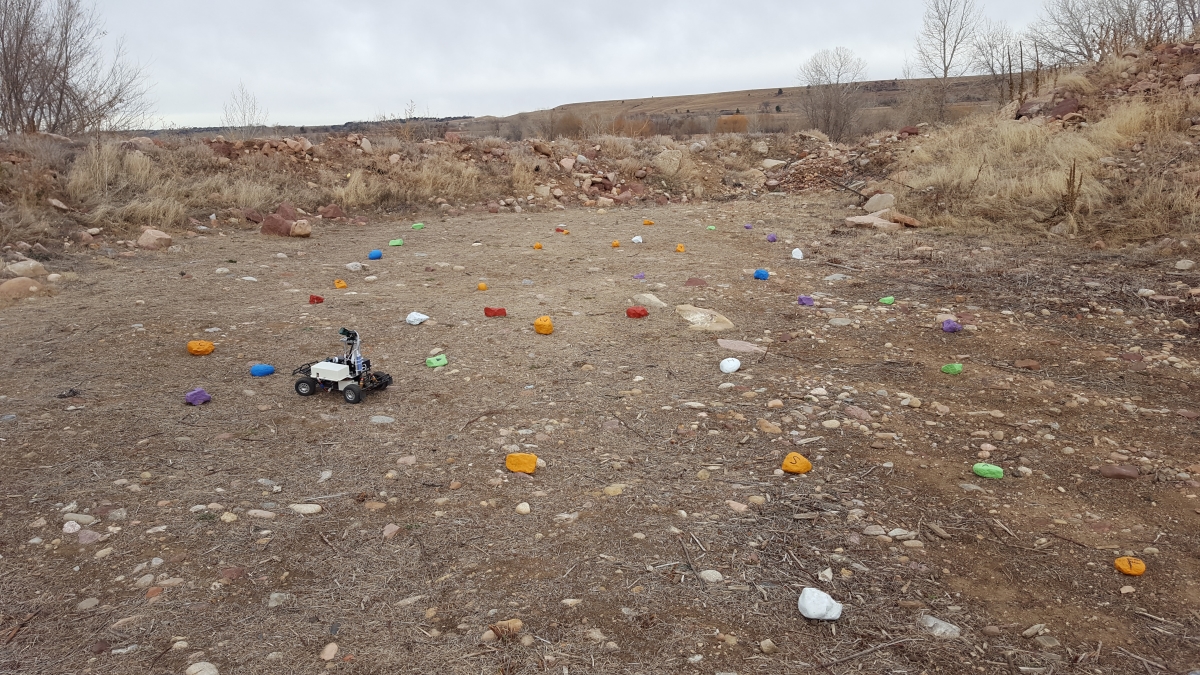}
\caption{Operator exploring the course on the University of Colorado-Boulder's South Campus via the rover. The course was surrounded by a rim of debris and rocks similar to a crater. Some of the painted targets the operator had to identify are seen scattered on the ground. \label{RoverCourse}}
\end{figure*}
Mars may be at the center of attention but it is not the sole benefactor for low-latency telerobotic missions. Before spending the immense time and money on a mission to Mars there should be smaller local tests, specifically on the lunar surface. Missions on the lunar surface will provide vast amounts of scientific discovery of previously unexplored areas on the Moon while also serving as a test platform to practice telerobotic procedures for future missions to Mars. The consequences of mistakes will be less severe on the lunar surface due to its proximity with Earth; therefore, telerobotic missions can take a step beyond just simulation and be put into use for lunar exploration. \par
The latency between Earth ground-stations and the lunar farside is approximately 2.6 seconds using a communication satellite at the Earth-Moon L2 point. Studies have shown that humans have a cognitive threshold for perceiving video as occurring in real-time for round-trip latency values of 0.3-0.4 seconds or less \cite{cogThreshold}. As a result, it is likely low-latency telerobotics will have an improvement over ground-based operations even when applied to bodies relatively close to Earth. The first crewed flights of the Orion will occur in the beginning of the next decade. To take advantage of these first flights telerobotic training could be paired with the installation of a low-frequency radio array on the farside of the Moon to peer at the early stages of the universe \cite{NESS}.\par
While low-latency telerobotics currently demonstrate its usefulness in applications on Earth in the fields of medicine and construction, it is time to apply low-latency surface telerobotics in the realm of space exploration in order to begin a new age of efficient scientific exploration of our solar system and beyond.

\subsection{Limitations of Telerobotics}
The benefits that occur when utilizing low-latency telerobotics for exploration in our solar system have been well researched and appear to be a promising option for scientific exploration \cite{HERRO}. However, the amount of research investigating the potential limitations of low-latency telerobotics has been modest to date (e.g. \cite{ISS}).\par

The limitations of telepresence for scientific exploration must be fully understood prior to implementing low-latency telerobotics as a strategy for exploring our solar system. Simulations of telerobotic use in space should be conducted from Earth to pinpoint the constraints of telerobotic operations. Some of the constraints might include the operational light level, required bandwidth, user interface, etc.\par
The maximum bandwidth available for communications between Earth-Moon L2 and the lunar farside is approximately 4 Megabits per second (Mbps) (assuming a 0.5 meter Ka-band antenna on the rover with 10 W of power) \cite{ISS}. However, the actual bandwidth available will sometimes drop below the 4 Mbps due to variable line-of-sight between the rover and the astronaut's communication antenna.\par

The first iteration of our experiment was brief and provided an overview of the effect of frames per second (FPS) on exploration of an unfamiliar environment. This overview allowed us to construct an experiment that focused our efforts of research in the correct FPS region to find meaningful results. In particular, we found that the minimum frame rate was most likely 4, 5, or 6 FPS.

\subsection{The Experiment}

We designed an experiment to address one factor that contributes to the application of telerobotics. 
In this experiment, we investigated the effects of lowered bandwidth, and in particular, the minimum operational FPS required for a human operator to successfully explore an unfamiliar environment using low-latency surface telerobotics.\par
To begin identifying the operational limits associated with low-latency telerobotic exploration our research group developed a TSS and an experiment to investigate limitations in exploration due to reduced video frame rate. The experiment consisted of volunteers remotely operating a rover in search of ``interesting" objects (targets). The operators searched for the targets using only the video stream feedback provided from the rover's two cameras. The frame rate was randomly varied for each trial and the time to discovery was used as the metric of success. This experiment was designed to identify the minimum frame rate required for a human to effectively explore an unfamiliar environment using low-latency telerobotics. The course and rover can be seen in Fig. \ref{RoverCourse}.
%%%%%%%%%%%%%%%% BANDWIDTH CALCULATION %%%%%%%%%%%%%%%
\section{Calculating Bandwidth from Frame Rate}
Frame rate is an abstraction of data rate or bandwidth. Finding the minimum frame rate will ultimately produce a minimum video/visual bandwidth necessary for effective exploration. The following is an equation to calculate the video bandwidth based on FPS, resolution, and colorscale:
\begin{equation*}
    \frac{\cancel{frames}}{second} \cdot \frac{\cancel{pixels}}{\cancel{frame}} \cdot \frac{bits}{\cancel{pixel}} = \frac{bits}{second}
\end{equation*}
The first term is how many frames are sent every second. The second term is the number of pixels in each frame; this affects the resolution of the image. The last term is the colorscale which corresponds to the number of color/shade combinations possible for each pixel. After some of the terms cancel we are left with the absolute worst case bits per second for a given set of parameters. This is the worst case because no compression strategies are taken into account.\par

Using the parameters from our experiment yields:
\begin{equation*}
    \frac{5\ \cancel{frames}}{second} \cdot
    \frac{(640\times480)\ \cancel{ pixels}}{\cancel{frame}} \cdot
    \frac{24\ bits}{\cancel{pixel}} = \frac{36.864\cdot 10^6\ bits}{second}\\
\end{equation*}
\begin{equation*}
    = 36.864\ Mbps
\end{equation*}
So, at worst our video bandwidth would be $36.864$ Mbps for one video stream. This was not our effective bandwidth though. Our video stream used a popular video compression codec called H.264. This codec can compress video by approximately 70\% to 93\%, depending on the motion present in the video \cite{h264}.\par

As mentioned previously, an estimate on the maximum bandwidth available for communications between Earth-Moon L2 and the lunar far-side is approximately 4 Mbps \cite{ISS}. The data rate calculated above does not fall under the 4 Mbps limit; however, by taking into account various strategies for reducing bandwidth our 36 Mbps can be drastically reduced and brought within the threshold for lunar far-side communications.\par
The first strategy is video compression. As mentioned above, H.264, has compression ranging from 70\% to 93\%. The second strategy decreases the colorscale by using a technique that compresses a 24 bit colorscale to an 8 bit colorscale without sacrificing quality \cite{trueColor}. Using these compression factors our video bandwidth would change from 36.864 Mbps to 1.72 Mbps. \par
Besides compression techniques, the actual maximum bandwidth available between Earth-Moon L2 and the lunar farside could be increased by either increasing the size of the antenna stationed at L2 or increasing the power of the signal sent from the rover.

%%%%%%%%%%%%%     MATERIALS AND METHODS     %%%%%%%%%%%%%%
\section{Materials and Methods}

\subsection{Experimental Method}
In this experiment we had operators identify targets that were randomly distributed within a lunar-analog course and measured the time it took to find a specified target. The targets were painted rocks with different symbols on the surface. The identification of a unique target was defined as a trial in this experiment.\par
\begin{figure}[h]
\centering
\includegraphics[width=0.48\textwidth]{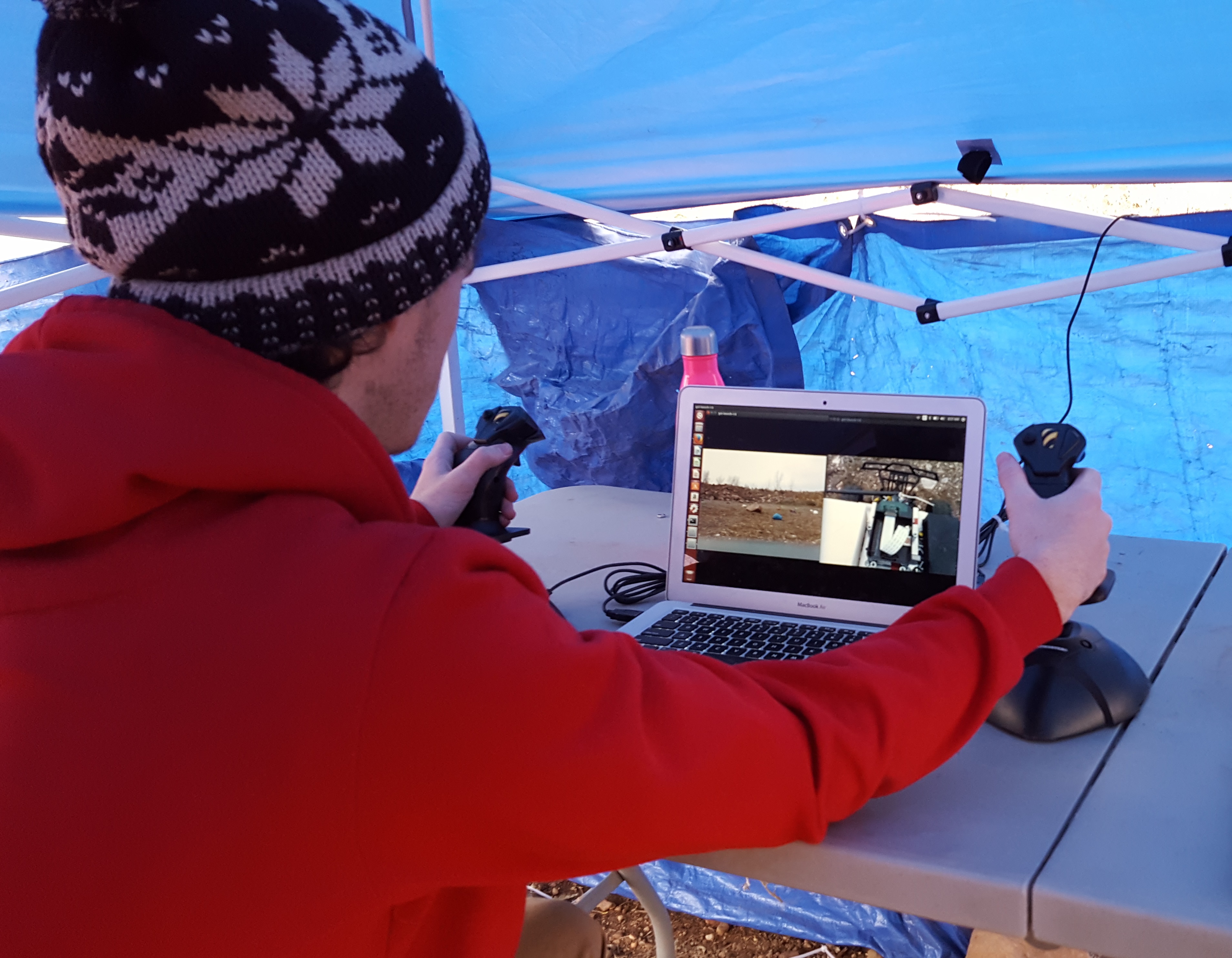}
\caption{Operator is housed inside a three sided tent to obstruct their view of the course. The two joysticks used to manipulate the rover are on either side of the computer. The left joystick was used to control the top camera and the right joystick was used to control the rover. There are two video feeds displayed: one is from the forward facing camera, the other is from the top mounted camera.
\label{Operator}}
\end{figure}

We had three different operators running the experiment and each operator ran 54 trials. Each trial corresponded to a search/exploration of a different target. The operators were isolated within an enclosed canopy to ensure they were operating the rover ``blind." Within this isolated environment, the operators controlled the rover through a computer interface. The two video feeds sent from the rover were displayed on the computer screen. The operator used two joysticks to control the rover's movement and the orientation of its top camera. Fig. \ref{Operator} shows an operator in the closed environment controlling the rover.\par

Before the operators began the official trials for data collection they underwent training by operating the rover in search of targets in a separate location. We repeated this process with each driver until the time to discovery became constant. This was done so that the learning effects present for each driver would be minimized before actual trials began. We also repeated the training process briefly at the start of a new testing day for each operator. This was done to ensure the operator was performing at the same level for each day of testing.\par
\begin{figure}[h]
\centering
\includegraphics[width=0.48\textwidth]{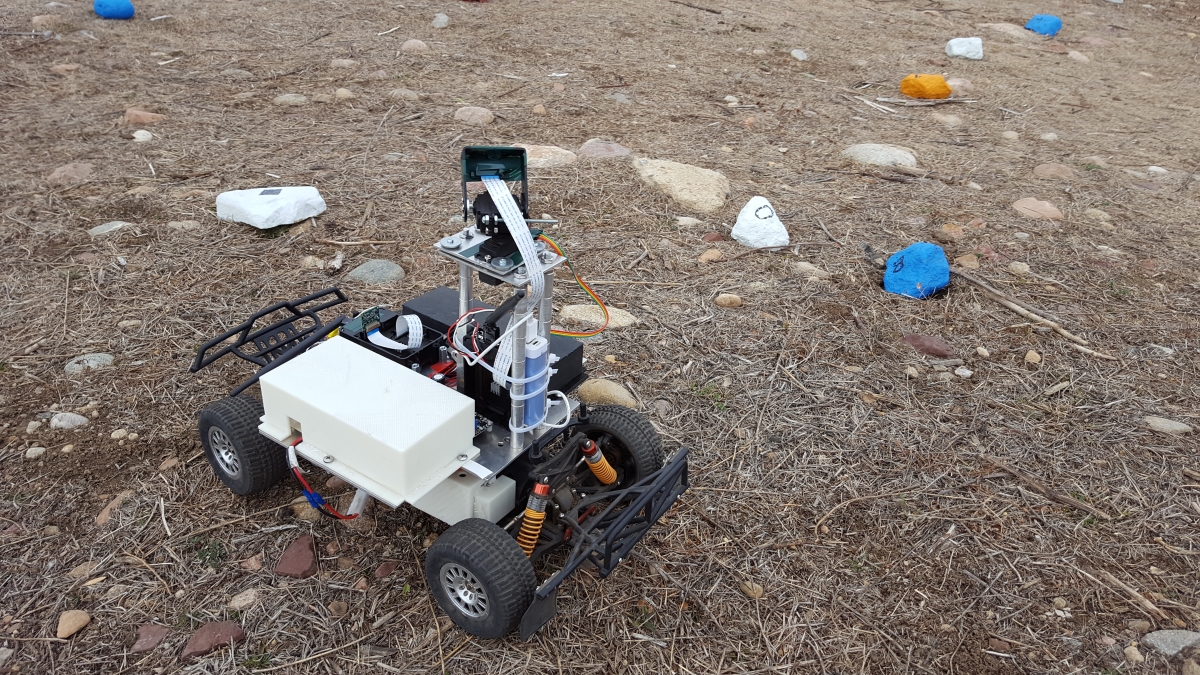}
\caption{Close up view of the rover. The forward facing stationary camera is mounted at the front and the top camera is mounted on two servos for manipulation. The various electronics controlling the rover are housed underneath the white and black protective shields. \label{RoverID}}
\end{figure}

The rover was located at the same starting point for each trial. Before initiating the trial we told the operator which target they needed to discover. While there were three of each type of targets scattered in the course, the operator only had to discover one of the three targets. The timer for each trial began as soon as the operator started moving the rover. The operator explored the course in search of the target, and the timer ended after successful identification of the target object. Fig. \ref{RoverID} shows the perspective from the rover when identifying a target.

\subsection{Rover}
The rover used was a modified remote-controlled car. Two new communication systems were installed; the first used a wireless router for streaming real-time video and the second used an XBee radio-frequency communication module for sending user commands. We used two Raspberry Pi's and Raspberry Pi cameras to capture real-time video. To stream video from the rover to the computer we used a video service called Gstreamer \cite{gstreamer}. This service allowed us to set the frame rate and aspect ratio of the video. The wireless router used the Transmission Control Protocol (TCP) to send data packets \cite{TCP}. Using TCP ensured that data packets were received in order and without error.\par
\begin{figure}[h]
\centering
\includegraphics[width=0.35\textwidth]{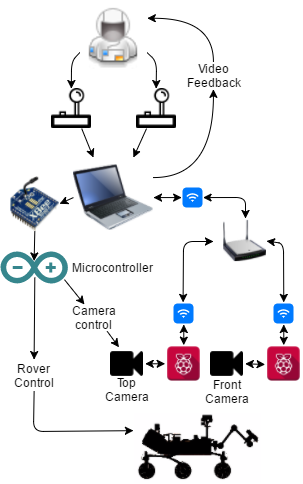}
\caption{The user interfaces with the command computer via two joysticks. Both joystick controls are sent via XBee to a micro-controller onboard the rover. One joystick manipulates the top camera and the other manipulates the rover. Two Raspberry Pi's onboard the rover send the video feeds via WiFi and WiFi router to the command computer to be displayed for the user. \label{diagram}}
\end{figure}

The XBee was connected to the rover with its mate connected to the operator's computer. The computer used two joysticks and a python script to create data packets that were sent over the computer's USB port to the XBee. The data packets sent to the computer's XBee were transmitted and then received by the XBee on the rover. Each data packet contained instructions to control the rover and its top camera. A micro-controller located on the rover parsed the incoming data packets and issued the instructions to the correct rover peripherals. Figure \ref{diagram} shows the flow of communications and control of the TSS.

\subsection{Targets for Identification} 
The targets for identification were painted rocks with the following colors: red, green, orange, blue, purple, or white. These colors served as the broad identifier for the operator while exploring. Each object also had a symbol written on the surface in bold black lettering. The symbols were chosen from a list of letters and shapes. The following were the symbols used: `A', `B', `C', `F', `I', `J', `M', `O', `S', `X', `Square', and `Triangle'. All of the combinations of colors and symbols yielded a total of 72 unique rocks. We made 3 of each unique rock type used in this experiment making a total of 216 rocks.

\subsection{Course}
The location used for running the experiment was located on the University of Colorado - Boulder's South Campus. It was chosen for its similarity with the Moon. The area of exploration used for the experiment was within the walls of a crater-like region on South Campus. The crater roughly formed an L shape. One end of the course had very rocky terrain while the rest of the course had minimal to moderate rocky terrain. The entire crater was sectioned off into approximately 180 five foot by five foot squares using twine and stakes. Once the the entire crater was sectioned into a grid each of the 216 rocks were randomly assigned a grid location. Since there were more rocks than grid spaces some of the grid locations contained more than one rock. This entire process was done so the course had a random distribution of rocks. Once all rocks were distributed the grid was removed from the course to eliminate an unnatural reference point for the operator's situational awareness.

%%%%%%%%%%%%%     RESULTS     %%%%%%%%%%%%%%
\section{Results}
We sought to identify the minimum frame rate in which humans can successfully explore an unfamiliar environment using low-latency surface telerobotics. We selected the dependent variable as the ``ability to explore" and the corresponding metric as time to discovery. The treatment variable was frame rate and evaluated at the following three levels: 4, 5, and 6 FPS. \par
The type I error used for the following statistical analyses was 5\%. 
The type I error is the probability of an incorrect rejection of a true null hypothesis. The type I error is used to compare against the p-values throughout our analysis. A p-value is the calculated probability of incorrectly rejecting the null hypothesis. It is compared against our type I error to determine whether the null hypothesis is accepted or rejected. If the p-value is less than our type I error we reject the null hypothesis.

%%%%%%%%%%%%%%%%%%%%%% Testing Normality %%%%%%%%%%%%%%%%%%%%%%%%%
\subsection{Testing for Normality}
\begin{figure}[h]
\centering
\includegraphics[width=0.5\textwidth]{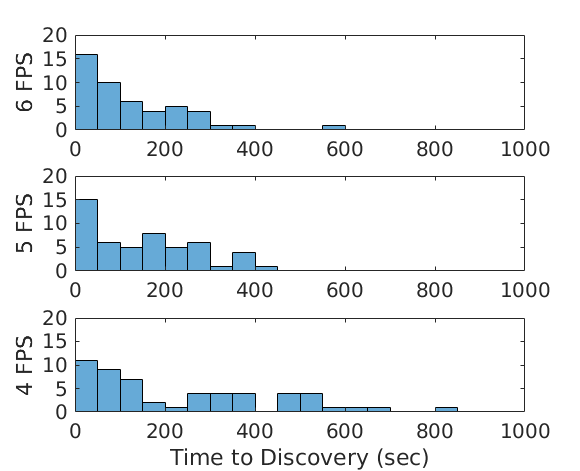}
\caption{Visually the dispersion of time to discovery increases as frame rate decreases. Also notice the long tails for each frame rates distribution that indicate a non-normal distribution. \label{histEachLevel}}
\end{figure}

The first assessment of the data consisted of a test for normality of the individual treatment levels of frame rate. The moment tests for skewness and kurtosis were performed in order to validate the underlying assumptions associated with the Analysis of Variance (ANOVA) and selection of the approach for statistical evaluation of homogeneity of variances. 
Fig. \ref{histEachLevel} shows the histograms of data by frame rate. The null hypothesis stated that the data are distributed normally, corresponding to values of Skewness ($\gamma_3$) and Kurtosis ($\gamma_4$) equal to zero. The alternative hypothesis stated that $\gamma_3$ or $\gamma_4$ do not equal zero, implying the data were not distributed normally. \par

Calculating $\gamma_3$ and $\gamma_4$ for the data within each level of frame rate using The Single-Sample Test for Evaluating Population Skewness and Kurtosis \cite{GaussianStatsTest} yields the results displayed in Table \ref{normalityTests}.

\begin{table}[h]
\centering
\textbf{\caption{Normality Tests\label{normalityTests}}}
\begin{adjustbox}{max width=0.48\textwidth}
\begin{tabular}{c|c|c|c|c|c}
Frame Rate & N & Skewness & p-value & Kurtosis & p-value \\ \hline
6 FPS & 48 & 1.545 & 1.574E-4 & 3.113 & $<$0.02 \\
5 FPS & 51 & 0.540 & 0.103 & -0.805 & $>$0.10 \\
4 FPS & 54 & 0.816 & 0.016 & -0.256 & $>$0.10    
\end{tabular}
\end{adjustbox}
\end{table}
\FloatBarrier

Based on the skewness and kurtosis p-values in Table \ref{normalityTests}, we reject the null hypothesis at the 95\% confidence level and will treat the data as non-normal for the homogeneity of variance analysis. Although the treatment level of 5 FPS passed the tests for normality, we must use the approach for non-normal data based on the rejection of the null hypothesis for both 6 FPS and 4 FPS.

%%%%%%%%%%%%%%%%%%%%%% VARIANCE ANALYSIS %%%%%%%%%%%%%%%%%%%%%%%%%%
\subsection{Homogeneity of Variance Analysis}
An assessment for homogeneity of variance, between the treatment levels of frame rate (4, 5, and 6 FPS), was performed in order to validate the underlying assumptions associated with the ANOVA and selection of the appropriate statistical approach for any subsequent post-hoc analyses of the means. Since the data are not normally distributed we used Levene's Improved Test for Homogeneity of Variances (Brown-Forsythe) using the Absolute Deviation from the Medians (ADM) \cite{LeveneImproved}. The null hypothesis stated that the variance between the treatment levels of frame rate were equivalent. The alternative hypothesis stated that the variance of at least one of the frame rate levels was not equivalent to the others.

\begin{figure}[h]
\centering
\includegraphics[width=0.48\textwidth]{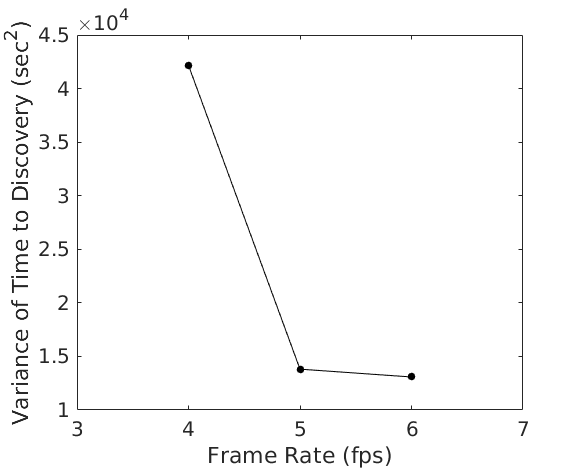}
\caption{Plot of sample variances. The contrast between 4 FPS and 5 FPS show how unreliable exploration is at 4 FPS compared to 5 FPS. Note: lines connecting data points are included to help guide the eye. \label{varAn_plot}}
\end{figure}

Table \ref{meansTable} lists the variance at each level of frame rate. Based on the results of the ANOVA using the ADM, the null hypothesis was rejected at the 95\% confidence level ($F(2,148)= 12.962, p = 0.000006$) and the alternative hypothesis was accepted. Fig. \ref{varAn_plot} shows the plot of the variance for each frame rate.

\begin{table}[h]
\centering
\textbf{\caption{Variance by Frame Rate\label{meansTable}}}
\begin{adjustbox}{max width=0.48\textwidth}
\begin{tabular}{c|c|c}
Frame Rate & N & Variance \\ \hline
6 FPS & 48 & 13080.936 \\ 
5 FPS & 51 & 13812.540 \\
4 FPS & 54 & 42164.710   
\end{tabular}
\end{adjustbox}
\end{table}
Since the variance was not equivalent between frame rates we conducted a post-hoc analysis to determine which groups were different from each other. We evaluated all pairwise comparisons using the Games-Howell test on the ADM due to unequal sample sizes in each treatment level of frame rate \cite{GamesHowell}. Table \ref{stuckVarPE} shows the results of the Games-Howell post-hoc test on the sample variances between all levels of frame rate. The sample variance of 5 and 6 FPS are equivalent at the 95\% confidence level, which forms Group 1 in Table \ref{stuckVarPE}. The sample variance of 4 FPS is significantly different than 5 and 6 FPS at the 95\% confidence level, which forms Group 2 in Table \ref{stuckVarPE}.\par
These results demonstrate that the consistency of exploration and discovery at 4 FPS has more variability than at 5 and 6 FPS. The consistency of exploration is a very important factor for future telerobotic missions given the cost and importance of mission success.

\begin{table}[h]
\centering
\textbf{\caption{Frame Rate Groups based on Variance \label{stuckVarPE}}}
\begin{adjustbox}{max width=0.48\textwidth}
\begin{tabular}{c|c|c|c}
Frame Rate & N & Group 1 & Group 2 \\ \hline
6 FPS & 48 & 13080.94 & \\
5 FPS & 51 & 13812.54 & \\ 
4 FPS & 54 & & 42164.71                        
\end{tabular}
\end{adjustbox}
\end{table}
\FloatBarrier

%%%%%%%%%%%%%%%%%%%%%%% Mean Time Analysis %%%%%%%%%%%%%%%%%%%%%%%%%%
\subsection{Mean Time to Discovery (MTD) Analysis}

The MTD was used as the main metric for measuring the operator's ``ability to explore" as the frame rate was varied. The null hypothesis stated that the MTD ($\mu$) across all frame rates would be equal. Our alternative hypothesis stated that the MTD of at least one frame rate was not equivalent to the others.\par
\begin{figure}[h]
\centering
\includegraphics[width=0.48\textwidth]{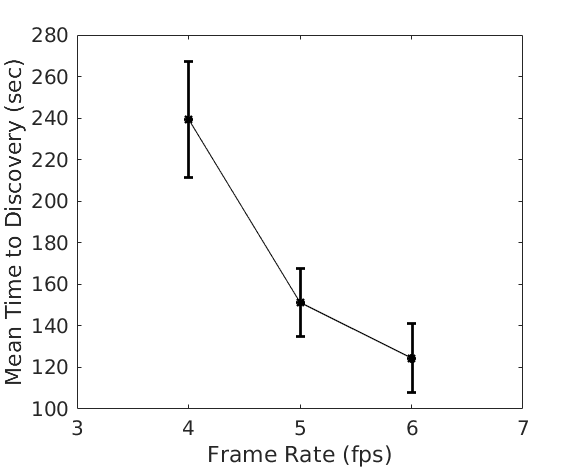}
\caption{The MTD has a significant decrease moving from 4 FPS to 5 FPS while moving from 5 FPS to 6 FPS has minimal effect and no statistical difference. This shows a type of cutoff point moving from 5 FPS to 6 FPS. Note: vertical lines represent error bars and lines connecting data points are included to help guide the eye. \label{margMeanPlot}}
\end{figure}

Table \ref{margMeanTable} lists the MTD calculated for each frame rate. Based on the results of the ANOVA using the MTD, the null hypothesis is rejected at the 95\% confidence level ($F(2,148)= 7.945, p = 0.000528$) and the alternative hypothesis is accepted. Fig. \ref{margMeanPlot} shows the plot of estimated MTD against frame rate. Table \ref{margMeanTable} lists the values of the points shown in Fig. \ref{margMeanPlot}. As shown in Fig. \ref{margMeanPlot} the MTD increases with a decrease in frame rate. The large difference between 5 and 4 FPS show an accelerated deterioration of MTD of exploration in an unfamiliar environment.

\begin{table}[h]
\centering
\textbf{\caption{Mean Time to Discovery \label{margMeanTable}}}
\begin{adjustbox}{max width=0.48\textwidth}
\begin{tabular}{c|c|c|c|c|c}
Frame & & & & & \\
Rate & N & MTD & Std. Deviation & Low & High \\ \hline
6 FPS & 48 & 124.5208 & 114.37192 & 2.00 & 554.00 \\
5 FPS & 51 & 151.3137 & 117.52676 & 7.00 & 410.00 \\
4 FPS & 54 & 239.5370 & 205.34046 & 17.00 & 819.00    
\end{tabular}
\end{adjustbox}
\end{table}

Since the MTD were not equivalent between frame rate levels, we conducted a post-hoc analysis to determine which groups were different from each other. We evaluated all pairwise comparisons using the Games-Howell test on the ADM due to unequal variances and unequal sample sizes in each treatment level of frame rate \cite{GamesHowell}. Table \ref{meanPointTable} shows the results of the Games-Howell post-hoc test on the MTD between all levels of frame rate. The MTD of 5 and 6 FPS are equivalent at the 95\% confidence level, which forms Group 1 in Table \ref{meanPointTable}. The MTD of 4 FPS is significantly higher than 5 and 6 FPS at the 95\% confidence level, which forms Group 2 in Table \ref{meanPointTable}. The results reinforce what we see in Figure \ref{margMeanPlot} and prove there is a significant increase in the MTD from 5 to 4 FPS. This rapid deterioration in the MTD signifies that a minimum of 5 FPS is necessary for effective exploration using our operation parameters.

\begin{table}[h]
\centering
\textbf{\caption{Frame Rate Groups based on MTD \label{meanPointTable}}}
\begin{adjustbox}{max width=0.48\textwidth}
\begin{tabular}{c|c|c|c}
Frame Rate & N & Group 1 & Group 2 \\ \hline 
6 FPS & 48 & 124.5208 & \\
5 FPS & 51 & 151.3137 & \\ %& \multirow{-2}{*}{137.917} \\
4 FPS & 54 & & 239.5370 %& 239.5370                    
\end{tabular}
\end{adjustbox}
\end{table}
\FloatBarrier

%%%%%%%%%%%%%     DISCUSSION     %%%%%%%%%%%%%%%
\section{Related Work \& Future Work}

We reviewed previous work on the effects of video quality to validate the results from our experiment. In particular, we examined studies that investigated the effects of video quality deterioration to determine a relationship between performance and video quality for video games. While the results from each of the studies differ slightly, ultimately the studies all concluded that there is a threshold value for video quality to maintain operability. \par
The first study \cite{videogame} explored the relationship between video frame rate and resolution on a user's ability to effectively shoot an opponent in a first-person-shooter video game. They came to the conclusion that user performance is up to 7 times worse at frame rates as low as 4 FPS compared to the performance when operating at 30 FPS. In particular, they found that the video game became essentially inoperable around 4 FPS.\par
The second study \cite{vidScaling} investigated the benefit of content-aware video stream scaling to adjust either the frame rate or resolution of the stream during periods of low bandwidth. The hypothesis stated that content-aware scaling would improve the apparent quality by performing the appropriate type of scaling for the video stream. Temporal scaling (dropping frames) occurred with low motion in the stream and quality scaling (reducing resolution) occurred with high motion in the stream. The study found that the use of content-aware scaling software can improve the apparent quality of the video stream by 50\%. \par
The last study \cite{Ranadive} inspected the effects of frame rate, resolution, and colorscale on an operators ability to perform a task. The tasks were specific to undersea teleoperation and included bolting/unbolting, lifting, opening/closing valves, connecting hoses, etc. The study determined that resolution and colorscale can be low while maintaining operability, while a frame rate below 5 FPS produces a considerable degradation in performance and increase in variability.\par
The results from these studies reinforce our data and conclusion. The second study indicates there is a relationship between minimum operational frame rate and the speed of the task. Future work should explore this relationship as it is relevant for future low-latency surface telerobotic missions, especially as telerobotic operation speeds increase. \par
In addition, future work should be invested into video compression applications for telerobotics. Applying powerful video compression to telerobotics in intelligent ways will significantly lower the necessary bandwidth for telerobotic operations; it may be the key to making telerobotics an extremely powerful and robust tool for future exploration. As we have described earlier, orbits around other planets will not always produce perfect bandwidth conditions. A high tolerance of bandwidth variability is crucial for effective and continuous teleoperation missions. Video compression is a technique that can significantly increase the tolerance of bandwidth variability by drastically decreasing the the worst case bandwidth required for telerobotic missions.\par

%%%%%%%%%%%%%     CONCLUSION     %%%%%%%%%%%%%%%
\section{Conclusion}

Our objective in this experiment was to investigate the effects of lowered bandwidth, in particular, to find the minimum operational frames per second (FPS) required for a human operator to successfully explore an unfamiliar environment using low-latency surface telerobotics. Our results show that as frame rate increases, time to discovery decreases slowly. Eventually a low enough frame rate causes a cutoff in operability and time to discovery increases quickly. There are many variables that determine the exact placement and shape of this curve. These variables include: FPS, resolution, colorscale, task performed, force-feedback, etc. Our data fit the trend that many other frame rate experiments produced and shows that exploring unfamiliar environments given our resolution, colorscale, and operation speed requires a minimum of 5 FPS.

%%%%%%%%%%%%%  Vitae %%%%%%%%%%%%%%%%%%%%
\section*{Vitae}
\begin{figure}[H]
\includegraphics[width=0.30\textwidth]{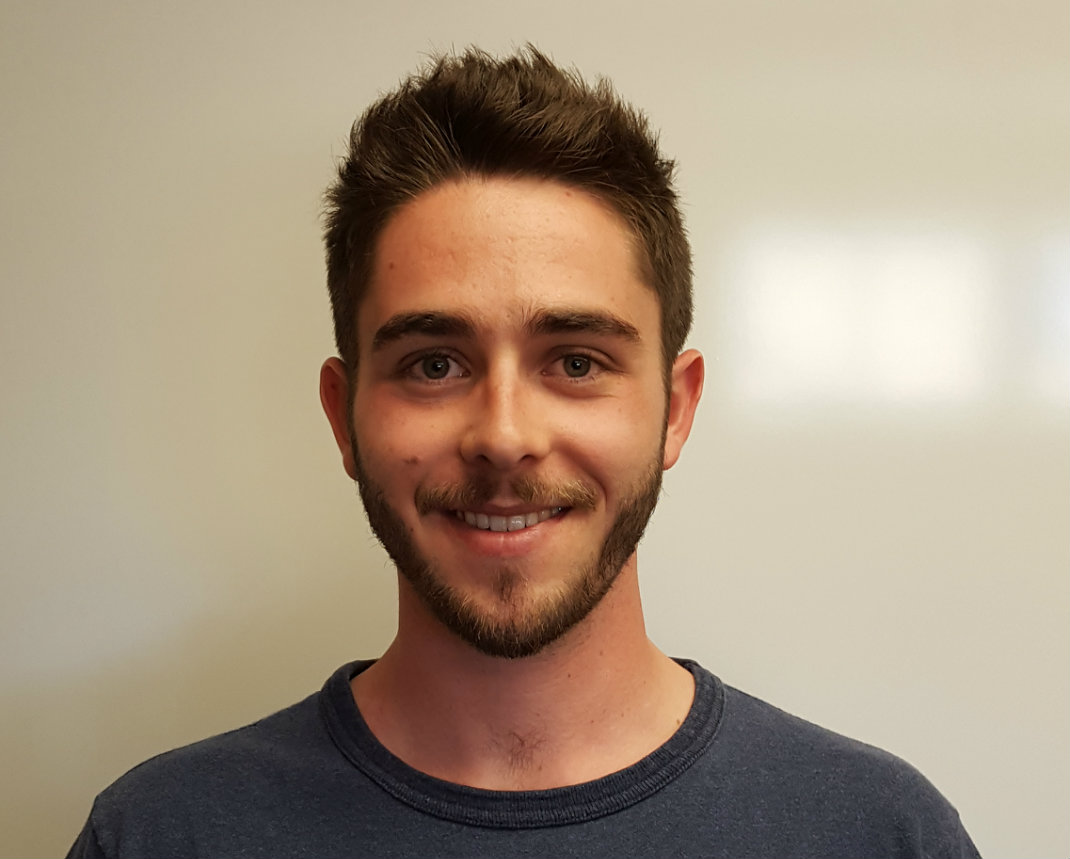} 
\label{Mellinkoff}
\end{figure}
\begin{itemize}
\item Benjamin Mellinkoff: Undergraduate studying Aerospace Engineering Sciences at the University of Colorado-Boulder. Manager of Center for Astrophysics and Space Astronomy's undergraduate telerobotics group.
\end{itemize}

\begin{figure}[H]
\includegraphics[width=0.30\textwidth]{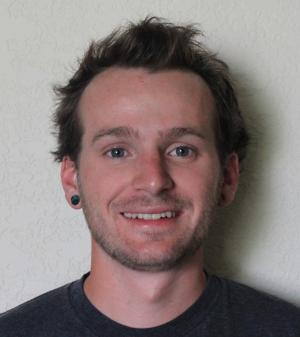} 
\label{Spydell}
\end{figure}
\begin{itemize}
\item Matthew Spydell: Undergraduate studying electrical engineering and computer engineering at the University of Colorado-Boulder. He is the electrical and computer engineer for the Center for Astrophysics and Space Astronomy's undergraduate telerobotics group.
\end{itemize}

\begin{figure}[H]
\includegraphics[width=0.30\textwidth]{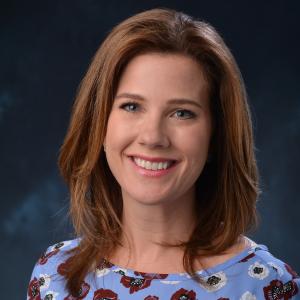} 
\label{Burns}
\end{figure}
\begin{itemize}
\item Wendy Bailey: Instructor in the Lockheed-Martin Engineering Management Program, teaching in the area of Quality Science. She earned her undergraduate degree in Mechanical Engineering from Purdue University, and a Masters of Engineering from the Lockheed-Martin Engineering Management Program at the University of Colorado Boulder, with an emphasis in six sigma, quality systems and applied statistics. Prior to her graduate degree, she was trained in statistical methods by Luftig \& Warren International (LWI). Wendy also worked for 14 years at Anheuser-Busch, where she became skilled in the application of statistics in an industrial environment.
\end{itemize}

\begin{figure}[H]
\includegraphics[width=0.30\textwidth]{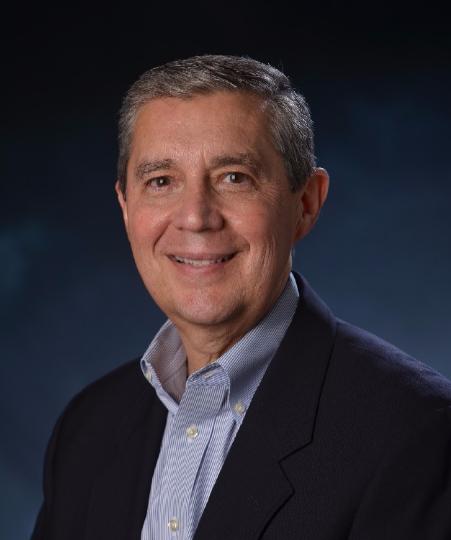} 
\label{Burns}
\end{figure}
\begin{itemize}
\item Dr. Jack Burns: Professor in the Department of Astrophysical and Planetary Sciences and Vice President Emeritus for the University of Colorado. He is also Director of the NASA-funded SSERVI Network for Exploration Space Science (NESS). Burns is an elected Fellow of the American Physical Society and the American Association for the Advancement of Science. He received NASA's Exceptional Public Service Medal in 2010 and NASA's Group Achievement Award for Surface Telerobotics in 2014. Burns was a member of the Presidential Transition Team for NASA in 2016/17. Burns serves as senior Vice President of the American Astronomical Society.
\end{itemize}
%%%%%%%%%%%% Acknowledgements %%%%%%%%%%%%%%%%%%
\section*{Acknowledgements}
Funding: This work was supported by the Lockheed Martin Space Systems Company; and the NASA Solar System Exploration Research Virtual Institute. We especially thank Chris Norman and Raul Monsalve for providing their counsel.

\bibliographystyle{elsarticle-num}

%\input{elsarticle-template.bbl}
%\bibliographystyle{elsarticle-num}
%\bibliography{biblio}
%\begin{thebibliography}{00}

\end{document}